\documentclass[twocolumn, amsmath,amssymb,aps, superscriptaddress]{revtex4-1}

\usepackage{verbatim}
\usepackage{graphicx}
\usepackage{bm} % bold math 
\usepackage{ifpdf} % This lets us get figures right both in pdf and ps versions
\usepackage{dcolumn}  % Align table columns on decimal point
\usepackage{epsfig}
\usepackage{color}
\usepackage[usenames,dvipsnames]{xcolor}
\usepackage{siunitx}

\def\rmd{{\mathrm{d}}}

\newcommand{\kT}{k_\mathrm{B}T}
\def\eq{Eq.}

\newcommand{\celsius}{\ensuremath{^\circ}C}

\begin{document}

\title{Surfactant Mediated Particle Aggregation in Nonpolar Solvent}
\author{Mojtaba Farrokhbin}
\affiliation{Department of Inorganic and Analytical Chemistry, University of Geneva,
Sciences II, 30 Quai Ernest-Ansermet, 1205 Geneva,
Switzerland}
\affiliation{Department of Physics, Faculty of Sciences, Yazd University, Yazd 89195-741, Iran}
\author{Biljana Stojimirovi\'c}
\affiliation{Department of Inorganic and Analytical Chemistry, University of Geneva,
Sciences II, 30 Quai Ernest-Ansermet, 1205 Geneva,
Switzerland}
\author{Marco Galli}
\affiliation{Department of Inorganic and Analytical Chemistry, University of Geneva,
Sciences II, 30 Quai Ernest-Ansermet, 1205 Geneva,
Switzerland}
\author{Mohsen Khajeh Aminian}
\affiliation{Department of Physics, Faculty of Sciences, Yazd University, Yazd 89195-741, Iran}
\author{Yannick Hallez}
\affiliation{Laboratoire de G\'enie Chimique, Universit\'e de Toulouse, CNRS, INPT, UPS, Toulouse, France}
\author{Gregor Trefalt}
\email{E-mail: \texttt{gregor.trefalt@unige.ch}}
\affiliation{Department of Inorganic and Analytical Chemistry, University of Geneva,
Sciences II, 30 Quai Ernest-Ansermet, 1205 Geneva,
Switzerland}

\date{\today}

\begin{abstract}
Aggregation behavior of particles in nonpolar medium is studied with time-resolved light scattering. At low concentrations of surfactant particles are weakly charged and suspensions are not stable. Suspensions get progressively more stable with increasing surfactant concentration as particles get more highly charged. At high concentrations the particles get neutralized and aggregation is again fast. The theory of Derjaguin, Landau, Verwey, and Overbeek (DLVO) is able to predict the stability ratios quantitatively by using the experimentally measured surface charge, screening lengths and van der Waals forces.

\end{abstract}

\maketitle

%%%%%%%%%%%%%%%%%%%%%
%%% Introduction %%%%
%%%%%%%%%%%%%%%%%%%%%

\section{Introduction}

Charging of particles in nonpolar media is important for many practical applications, for example, development of electrophoretic displays~\cite{Comiskey1998, Chen2003}, airborne drug delivery systems~\cite{Jones2006}, and toner technologies~\cite{Croucher1985}. In addition of being important for applications, nonpolar colloids are also extremely interesting from a fundamental point of view. The understanding of charging and stability mechanisms of particles in nonpolar media is far from being completely achieved. Furthermore, comparisons vis-a-vis aqueous systems could give new insights of possible charging mechanisms.

The generation of charges in solutions typically occurs by the mechanism of dissociation of salt. The important parameter in this context is the Bjerrum length, which is the distance between two elementary charges for which the electrostatic energy is equal to the thermal energy. For water, Bjerrum length is smaller than 1~nm and is comparable to the size of the hydrated ions, therefore the dissociation of salt is spontaneous. In nonplolar media with very low dielectric constant, such as alkanes, the Bjerrum length becomes large and can reach values of few tens of nanometers. In such solutions the ions stay paired and do not dissociate spontaneously. However, with the addition of ionic~\cite{Morrison1993, Smith2012, Hsu2005,Roberts2008,Sainis2008} or even non-ionic surfactants\cite{Guo2009, Poovarodom2010, Gacek2012, Lee2016, Espinosa2010}, charged species are formed, which can be observed by increased conductivity of surfactant solutions. These surfactants form inverse micelles, which are much bigger than simple ions. The charge is then formed by charge disproportionation, where two neutral micelles exchange a charge forming two oppositely charged species~\cite{Morrison1993,Hsu2005,Roberts2008,Sainis2008}. Another possibility is to use electrolytes with large, bulky ions, which also dissociate partly in nonpolar solvents~\cite{Waggett2018, Finlayson2016a}.

When solid surfaces or colloidal particles are immersed in nonpolar solvents they typically do not get charged. However, surface charge can be acquired by the introduction of surfactants. Surface charging mechanisms in nonpolar liquids mediated by surfactants was studied with various methods~\cite{Morrison1993, Smith2012}. Both the nature of the surface and the properties of the surfactants are important in this process. Different charging mechanisms have been described. The charge can form by dissolution of surface ions into reverse micelles~\cite{Briscoe2002a}, preferential adsorption of charged micelles onto the surface~\cite{Morrison1993, Roberts2008}, or adsorption of individual surfactant molecules and their subsequent dissociation~\cite{Smith2006, Kemp2010, McNamee2004}. In the case of non-ionic surfactants the charging can be explained by acid-base mechanisms~\cite{Morrison1993, Smith2012, Espinosa2010, Guo2013, Lee2016, Poovarodom2010, Gacek2012, Gacek2012a}. Specifically designed particles can charge also without addition of surfactant by surface dissociation of bulky ionic liquid like ions~\cite{Hussain2013, Sanchez2011}. 

Charging of colloids in nonpolar liquids leads to electrostatic repulsive forces between them. These repulsive forces were measured by different techniques, such as surface force apparatus~\cite{Briscoe2002a}, colloidal probe technique~\cite{McNamee2004}, total internal reflection microscopy~\cite{Prieve2008}, optical tweezers~\cite{Sainis2007,Sainis2008, Merrill2009, Merrill2010, Masri2011}, and pair correlation function measurements~\cite{Hsu2005, Espinosa2010}. The majority of these experiments measured a long-ranged tail of the electrostatic  interaction, and interpreted the measured profiles with screened Coulomb, also known as Yukawa potential. The parameters entering the Yukawa interaction extracted from the forces were found to be consistent with conductivity and electrokinetic measurements~\cite{Hsu2005, Espinosa2010, Sainis2008}, albeit some recent force measurements showed deviations of Debye screening lengths from values extracted from conductivity measurements~\cite{Waggett2018}.

The particle interactions drive the stability of colloidal particles in suspensions. Typically colloids are not stable in pure nonpolar liquids, but suspension get stabilized upon addition of surfactant. While stability in nonpolar suspensions have been observed qualitatively~\cite{Hsu2005, Espinosa2010, Smith2017b}, no quantitative data on aggregation kinetics is available. Such data would first quantify the stability of nonpolar suspensions and second enable to extract further information about the electrostatic and van der Waals interactions in these systems. For aqueous suspensions the simultaneous measurements of aggregation kinetics and particle interactions on the same particles showed, that the interaction forces extracted from the colloidal probe technique can be used to evaluate quantitatively stability ratios~\cite{Sinha2013, MontesRuiz-Cabello2013, MontesRuiz-Cabello2015}. Conversely, information about the interactions, such as surface charge regulation properties, can be extracted from the aggregation rate measurements~\cite{Cao2017, Cao2018, Rouster2017, Rouster2019}.

Here we present the measurements of aggregation rates for three different particle suspensions in decane in the presence of a surfactant. In combination with the results from direct-force measurements and electrophoresis we were able to elucidate the aggregation mechanisms in nonpolar suspensions. We further pinpointed the most important factors affecting the aggregation process and extracted some information about surface charge regulation effects.

\section{Experimental Methods}

\subsection{Materials}

Aqueous suspensions of silica particles (5.0 wt.\%, Corpuscular Inc) and two types of surfactant-free polystyrene latex particles, namely sulfate latex (SL) (8.0~wt.\%) and amidine latex (AL) (4.0~wt.\%) (both from Invitrogen Corporation), were used to study the aggregation mechanism in nonpolar media. The sign of the surface charge of these particles in water is dictated by their surface chemistry. Silica and SL particles are negatively charged in water, while AL is positively charged. The average particle size determined and polydisperisty measured by the manufacturer with transmission electron microscopy (TEM) and dynamic light scattering (DLS) are given in Table~\ref{tab:particles}.
%%%%%%%%%%%%%%%%%%%%%%%%%%%%%%%%%%%%%%%%%%%
\begin{table*}[t]
	\caption{ Properties of the colloidal particles used in the experiments.} 
	\centering 
	\begin{tabular}{c c c c c c} 
		\hline 
        Particle  &\multicolumn{2}{c}{Radius (nm)} &Polydispersity Index$^a$ & Fast Rate in Water$^c$ & Fast Rate in Decane \\
                 & TEM$^a$       & DLS$^b$               & CV (\%) & $k_{\rm fast}$~($\times 10^{-18}$~m$^3$/s) & $k_{\rm fast}$~($\times 10^{-18}$~m$^3$/s)  \\ 
		\hline 
		Silica             & 100 & 103 & 22& $1.0\pm 0.2$ & $1.0\pm 0.2^d$ \\ 
		Sulfate Latex (SL) & 150 & 155 & 4.7 & $2.8\pm 0.3$ &$2.3\pm0.3^e$\\
		Amidine Latex (AL) & 110 & 117 & 4.3 & $2.8\pm 0.3$ & $2.3\pm 0.2^f$ \\
		\hline
		\multicolumn{6}{l}{\footnotesize $^a$Measured by the producer with electron microscopy, except for silica which was measured with DLS.}\\
        \multicolumn{4}{l}{\footnotesize $^b$Measured by dynamic light scattering in water at 25~\celsius .}\\
        \multicolumn{4}{l}{\footnotesize $^c$Measured at concentrations of KCl above 600 mM.}\\
        \multicolumn{4}{l}{\footnotesize $^d$Measured in 0.002-0.01 mM AOT concentration range.}\\
        \multicolumn{4}{l}{\footnotesize $^e$Measured in 0.02-0.4 mM AOT concentration range.}\\
        \multicolumn{4}{l}{\footnotesize $^f$Measured in 0.3-0.5 mM AOT concentration range.}
	\end{tabular}
	\label{tab:particles}
\end{table*}
%%%%%%%%%%%%%%%%%%%%%%%%%%%%%%%%%%%%%%%%%%%
Additional DLS measurements were performed in house for all three samples. The average size measured by DLS is slightly larger than the values obtained by TEM, probably due to polydispersity effects. The supplied polystyrene particle suspensions were first dialyzed in cellulose ester (SL particles) or polyvinylidene fluoride membranes (AL particles) against milli-Q water (Millipore) for about one week until the conductivity reached about $ 70~\mu$S/m. After dialysis the particle concentration of the suspension was determined with static light scattering by comparing the scattered intensity of dialyzed suspensions and non-dialyzed suspensions with known concentrations.

In order to study charging and aggregation in nonpolar media, the particles from aqueous suspensions had to be transferred into decane, which was achieved in two steps. First, small amount of dialyzed particle suspensions were injected into isopropyl alcohol (99.5~\%, Sigma-Aldrich) resulting in suspensions with concentrations of 5~g/L and 0.4~g/L for silica and polystyrene particles, respectively. The isopropanol suspensions were used in the second step to prepare samples in decane. Dioctyl sodium sulfosuccinate with the commercial name Aerosol-OT (AOT) ($ > 95\% $, Fisher Chemical) was dissolved in decane (Sigma-Aldrich, $\geq 99\%$) and filtered through $0.1~\mu$m  syringe filters. Finally, isopropanol suspensions were injected into decane/AOT reaching the final particle concentrations of 100~mg/L and 10~mg/L for silica and polystyrene particles, respectively. Note that the amount of isopropanol in the final suspensions was typically below 1 vol.\% and therefore these small amounts do not significantly change the properties of the solvent.

\subsection{Electrophoresis}
Phase analysis light scattering (PALS) was used to measure electrophoretic mobility on a Zetasizer Nano ZS instrument (Malvern). A dip cell ZEN 1002 appropriate for non-aqueous solutions was used. This cell employs two planar palladium electrodes separated by 2~mm, dipped into a glass cuvette with square cross-section. The cell electrodes and glass cuvettes were cleaned and rinsed with decane. Suspensions were prepared by mixing the stock solution of AOT surfactant in decane to get the desired surfactant concentration and then particles were injected from stock isopropanol suspensions. The final particle concentration used for electrophoresis measurement was $100$~mg/L for silica and $10$~mg/L for polystyrene particles, respectively. The mobility values were measured at different electric fields and extrapolated to zero-field values~\cite{Espinosa2010}. In general, the deviations of zero-field extrapolated values and values measured at finite field were below $5\cdot10^{-11}$~m$^2$/V/s, which corresponds to deviation in the electrokinetic potential below 5~mV. The electrophoretic mobilities were converted into electrokinetic potentials by employing the H\"uckel expression suitable for the systems under investigation.

\subsection{Karl Fischer titration}
Karl Fischer titration (736 GP Titrino, Metrohm) with methanol as the solvent was used to measure the water content of decane/AOT suspensions. The water content depended on the AOT concentration and it was in the range of 0.1--0.4~wt.\% for silica suspensions and 0.05--0.1~wt.\% for polystyrene suspensions. The higher water content for silica samples can be explained by the larger concentration of silica particles, which introduces more water into the final decane suspension.

\subsection{Conductivity Measurements}
High Precision Conductivity Meter (Model 1154, emcee electronics, inc) was used to measure the conductivity of the suspensions. Conductivities below 1~pS/m were measured for the pure solvent (decane). The conductivity values were used to estimate the inverse Debye length, $\kappa$.  

\subsection{Force Measurements}
The van der Waals forces between spherical silica particles (Bangs Laboratories Inc.) in decane were measured using the colloidal probe technique based on atomic force microscopy (AFM) in the symmetric sphere-sphere geometry. First, a single silica particle with diameter reported 4.07~$\mu$m was glued on a tipless cantilever (MicroMasch, Tallin, Estonia). A small drop of glue (Araldite 2000+) and some silica particles were placed on a glass slide next to each other. The cantilever was mounted inside the AFM head and manipulated to touch the glue and immediately after to pick up a single particle. The substrate was prepared separately by spreading silica particles on a quartz microscope slide (Plano GmbH, Wetzlar, Germany), previously cleaned in piranha solution (3:1 mixture of H$_2$SO$_4$ (98\%) and H$_2$O$_2$ (30\%)). Both cantilever with a glued particle and prepared substrate were placed in an oven for 2~h at 1200~\celsius . After this sintering process, the particles were firmly attached to the substrate/cantilever and the glue was burned away.

Force measurements were done at room temperature $23\pm 2$~\celsius\ with a closed loop AFM (MFP-3D, Asylum Research) mounted on an inverted optical microscope (Olympus IX70). The substrate and the cantilever were cleaned in ethanol and water, and plasma treated for 20~min. The substrate with particles was mounted into the fluid cell. Decane (Acros Organics, 99\%) was kept on molecular sieves to reduce water content, and filtered through 0.02~$\mu$m syringe filter (Whatman Anotop 25) before measurement.  The particle on the cantilever was centered above the selected particle on the substrate with a precision of about 100~nm. The deflection of the cantilever was recorded for 150 approach-retract cycles, and the cantilever speed was 400~nm/s. The deflection was converted to force using Hooke's law, where the spring constant of the cantilever was determined by the Sader method, and was 0.345~N/m. The approach part of the recorded curves was averaged to obtain final force-curves.

\subsection{Light Scattering}

A time-resolved light scattering technique was used to study the particle aggregation mechanism by using a goniometer setup (ALV/CGS-3). This instrument uses He/Ne laser with a wavelength of 633~nm. Stock solution of AOT surfactant in decane was diluted to the desired concentration in 2~mL borosilicate cuvettes. Then, an appropriate amount of suspension of particles in isopropanol was injected into the cuvette and the suspension was rapidly mixed. Final concentrations of particles in the samples were $100$~mg/L for silica and $10$~mg/L for polystyrene particles, respectively. A higher concentration of silica particles had to be used due to their lower refractive index, which results in lower scattering signal. Before use, borosilicate cuvettes were cleaned in a hot piranha solution for one day, then washed with Milli-Q water and dried in a dust-free oven at 60~\celsius. The scattering intensity was accumulated at an angle of 90\ensuremath{^\circ} for $20$~s to build a correlation function, which was analyzed with a second-order cumulant fit in order to obtain the diffusion coefficient. The Stokes-Einstein relation was used for the conversion of the diffusion coefficient to hydrodynamic radius, where the viscosity of pure decane of 0.86~mPas was used.

In order to determine the aggregation rate constants for singlet particles forming a doublet, the time evolution of the hydrodynamic radius was followed for typically 1~h~\cite{Holthoff1996}. The apparent dynamic rate coefficient, $\Delta$, was determined from the initial increase of the hydrodynamic radius, $R_{\rm h}$,
\begin{equation}
\Delta = \left . \frac{1}{R_{\rm h}(0)}\cdot\frac{\rmd R_{\rm h}(t)}{\rmd t} \right |_{t\to 0} .
\label{eq:app_dyn_rate}
\end{equation}
The stability ratio, $W$, was determined as
\begin{equation}
W = \frac{\Delta_{\rm fast}} {\Delta} ,
\label{eq:stab_ratio}
\end{equation}
where $\Delta_{\rm fast}$ represents the fast apparent dynamic rate, determined at conditions, where attractive van der Waals forces dominate the interparticle interaction. The fast absolute aggregation rates, $k_{\rm fast}$,  were calculated by using expression~\cite{Holthoff1996}
\begin{equation}
\Delta_{\rm fast} = \frac{I_2(q)}{2I_1(q)}\left(1- \frac{1}{\alpha} \right)N_0 k_{\rm fast} ,
\label{eq:app_rate}
\end{equation}
where $t$ is time, $q$ is the magnitude of the scattering vector, $I_2$ is the scattering intensity of a doublet, $I_1$ is the scattering intensity of a singlet, $\alpha = 1.38$ is the hydrodynamic factor, and $N_0$ is the initial number concentration of singlets. The hydrodynamic factor is the ratio of the effective hydrodynamic radius of the dublet and hydrodynamic radius of the singlet. The numerical value of 1.38 is calculated based on dublet diffusion coefficient in a low Reynolds number fluid~\cite{Holthoff1996}. The ratio of doublet and singlet scattering intensities can be calculated with the Rayleigh-Gans-Debye (RDG) theory~\cite{Holthoff1996},
\begin{equation}
\frac{I_2(q)}{2I_1(q)} = \frac{\sin(2qa)}{2qa} + 1,
\label{eq:app_rate}
\end{equation}
where $a$ is the radius of the particle. For the relevant conditions we have checked the accuracy of the RDG with more precise T-matrix theory~\cite{Galletto2005} based on Mie scattering. The results of the RDG calculations were within 2~\% of the T-matrix results.

\section{Results and Discussion}

The aggregation of colloids in decane in the presence of AOT surfactant was investigated. In order to understand the aggregation process, we have first measured the charging of silica, sulfate latex (SL), and amidine latex (AL) in decane as a function of AOT concentration. All the particles are practically uncharged in pure decane and get charged upon the addition of the surfactant. We were further able to estimate the interaction forces between the particles which in turn enabled us to calculate the aggregation rate constants. Comparison between the theoretically calculated and experimentally measured aggregation rates confirms that the electrostatic and vdW interactions are the main drivers of the aggregation process.

\subsection{Charging of Colloidal Particles}

\begin{figure}[t]
\centering
\includegraphics[width=8.5cm]{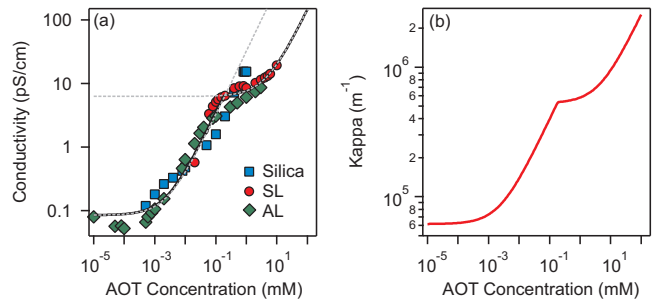}
\caption{(a) Conductivity and (b) inverse Debye length $\kappa$ versus concentration of AOT in decane solutions containing silica, SL, and AL particles. The dashed lines in (a) are linear fits for conductivity below and above CMC, the full line shows the interpolation fit. Note that the conductivity values obtained in solutions without particles are the same as in particle suspensions within experimental error. The contribution of colloids to conductivity is therefore negligible.}
\label{fig:conductivity}
\end{figure}

Charging of colloidal particles was determined indirectly by measuring electrophoretic mobility. Electrophoretic mobility can be further converted to electrokinetic potential and electrokinetic charge. This conversion depends on the thickness of the double-layer compared to the size of the particle. The double-layer thickness, which is equal to the inverse Debye length, $\kappa^{-1}$, is dependent on the concentration of charged species in the solution. The charge in the nonpolar solutions containing AOT, steams from inverse micelles. However, only a small fraction $\sim 1\cdot 10^{-5}$ of these micelles are charged~\cite{Hsu2005}. Electrical conductivity measurements in the nonplolar solutions give an accurate estimation of the Debye lengths~\cite{Hsu2005, Espinosa2010, Sainis2008}. By knowing the size of charge carriers $\kappa$ can be calculated from conductivity, $\sigma$, by using
\begin{equation}
\kappa = \sqrt{\frac{24\pi^2 \eta r_{\rm ion} \lambda_{\rm B} \sigma}{e_0^2}} ,
\label{eq:conductivity}
\end{equation}
where $\eta = 0.86$~mPas is the viscosity of decane, $r_{\rm ion}$ is the radius of ions, $\sigma$ is the conductivity, and $\lambda_{\rm B} = e_0^2/(4\pi \varepsilon_{\rm r}\varepsilon_0 \kT) = 28.0$~nm is the Bjerrum length, where $e_0$ is the elementary charge, $\varepsilon_{\rm r} = 2.0$ is the relative dielectric permittivity of decane, $\varepsilon_0$ is the vacuum dielectric permittivity, $k_{\rm B}$ is the Boltzmann constant, and $T$ is the absolute temperature. For the radius of the AOT inverse micelles a value of 2~nm was used~\cite{Hsu2005}. All values used in \eq~\ref{eq:conductivity} correspond to room temperature. Fig.~\ref{fig:conductivity}
%%%%%%%%%%%%%%%%%%%%%%%%%%%%%
%%%%%%%%%%%%%%%%%%%%%%%%%%%%%
shows the experimental values of conductivity and calculated inverse Debye length for samples containing silica, SL, and AL particles. The conductivity can be fitted by two linear fits, before and above the critical micelle concentration (CMC), respectively. The measured conductivity values are comparable to values reported in the literature~\cite{Hsu2005, Guo2009}. The CMC is located at $\sim 0.2$~mM, which is comparable to the value obtained earlier by SANS~\cite{Kotlarchyk1985}. Typical values of CMC for AOT in nonpolar solvents are found in the range 0.1--5~mM and depend on the water content and other impurities~\cite{Kotlarchyk1985, De1995}, in almost completely dry solvent the CMC value can be substantially smaller~\cite{Sainis2008, Kemp2010}.

Below the CMC the charge in solution is created by dissociation of individual AOT molecules, complexation of charged impurities by AOT, and formation of pre-micellar complexes~\cite{Guo2009, Sainis2008}. Above the CMC the conductivity is dominated by charged reverse micelles. The majority of micelles are neutral, however during collisions between them the charge can be transferred through charge disproportionation mechanism~\cite{Hsu2005, Guo2009, Sainis2008}, which can be described in the framework of charge fluctuation theory~\cite{Eicke1989a}.

The Debye lengths estimated from the conductivity are found in the range of $\kappa^{-1} = 1-10$~$\mu$m, see Fig.~\ref{fig:conductivity}b. Compared to the radius, $a$, of investigated colloids, which are between 0.10 and 0.15~$\mu$m, the double-layer thickness is much larger and therefore our measurements are done in the $\kappa a \ll 1$ limit.

\begin{figure}[t]
\centering
\includegraphics[width=8.5cm]{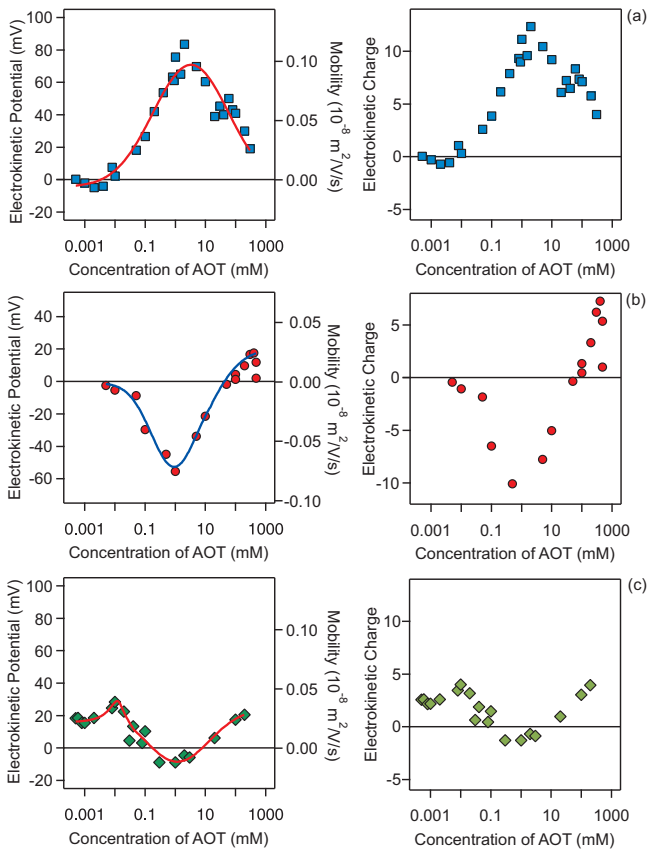}
\caption{Electrokinetic potential (left) and electrokinetic charge (right) of (a) silica, (b) SL, and (c) AL particles in decane as a function of AOT concentration. The right axis of the left panel represents the measured electrophoretic mobilities. The electrokinetic charge is in units of elementary charge. The lines in the left panel are empirical functions used for data interpolation.}
\label{fig:mobility}
\end{figure}

The electrophoresis results for all three systems are presented in Fig.~\ref{fig:mobility}.
%%%%%%%%%%%%%%%%%%%%%%%%%%%%%
%%%%%%%%%%%%%%%%%%%%%%%%%%%%%
The electrophoretic mobilities, $\mu$, were converted into the electrokinetic potential, $\zeta$, also called zeta potential, using the H\"uckel theory
\begin{equation}
\mu = \frac{2\varepsilon_{\rm r}\varepsilon_0 \zeta}{3\eta} ,
\label{eq:huckel}
\end{equation}
as is appropriate for the present dilute suspensions with $\kappa a \ll 1$.
The electrokinetic potential is further converted to the electrokinetic charge by~\cite{Hsu2005, Espinosa2010}
\begin{equation}
Z = \frac{a(1+\kappa a)}{\lambda_{\rm B}}\cdot\frac{\zeta e_0}{\kT} .
\label{eq:potential2charge}
\end{equation}
Note that this charge represents the number of charged species on the surface of one particle. The maximum absolute number of charges is relatively low as it reaches values of only about 10 charges per particle. However, due to low dielectric constant of the medium, this low charge translates into rather high maximal electrostatic potentials of about 80~mV. Eq. (\ref{eq:potential2charge}) is valid in fact only when the Debye-H\"uckel theory is valid, in practice for potentials lower than 40-50 mV in dilute suspensions. We use (\ref{eq:potential2charge}) however as $\zeta$ rarely exceeds this threshold in the present experiments. All the particles are only weakly charged at low surfactant concentration and the sign of the charge is consistent with their surface chemistry in water, namely negative for silica and SL (silanol and sulfate groups), and positive for AL (amidine groups). In all the systems charge reversal is observed upon addition of AOT. Note that the magnitude of the charge of the silica is extremely low at small AOT concentrations and therefore the particles are practically neutral at this conditions.

Electrokinetic potential of silica particles is almost neutral below 0.01 mM AOT and at this concentration silica surface becomes positively charged. After this point the potential increases and reaches a maximum value of about 80~mV at 3~mM of AOT. At high concentrations of AOT, mobility again decreases and approaches zero at concentrations of several hundreds mM. Similar behavior was observed for larger silica particles in decane/AOT suspensions by Keir et al.~\cite{Keir2002}. They have speculated that the charge reversal is caused by adsorption of positively charged AOT micelles. Such behavior could be also explained by specific adsorption of sodium ions to the silica surface.

Compared to the silica particles, the situation is reversed for the SL particles. The oposite behavior of silica and latex particles is possibly due to different affinity of AOT for the adsorption for the two surfaces. For SL particles the electrokinetic potential first decreases with concentration and reaches a pronounced minimum of $-55$~mV at 1~mM. At higher concentration the SL particles get again neutralized. This neutralization at higher concentrations is very reminiscent of the charging of SL particles in water~\cite{Oncsik2016, Cao2017, Sugimoto2018, Guo2013}. One can argue that the main driver for the charging in water and in decane/AOT is the charging of the sulfate groups present on the surface of the particle. In the case of decane the particle counterions can be solubilized by AOT micelles and this process would produce negatively charged sulfate groups on the surface~\cite{Guo2013}. However, looking at the electrokinetic charge profile in decane (Fig.~\ref{fig:mobility}, right), clear difference with aqueous suspensions is revealed. First, the number of surface charges on a similar particle in water is typically at least three orders of magnitude larger as compared to the values in nonpolar media. Second, while in the case of aqueous suspensions the charge of SL particles at higher concentrations is constant and the reduction in the magnitude of potential is caused by double-layer screening with salt, in decane/AOT system, the charge is not constant, and the magnitude of charge first decreases and after 1~mM increases. Therefore in the latter case it seems that there are two adsorption regimes which control the charge and consequently the potential. The behavior at lower concentration is consistent with mechanism of AOT adsorption and its subsequent ionization, which was inferred from scattering measurements for PMMA particles~\cite{Kemp2010}. At higher concentrations positive species are adsorbing, which could possibly be solubilized sodium ions.

The results for AL particles, shown in Fig.~\ref{fig:mobility}c, reveal that this type of surface remains weakly charged in the whole concentration regime, consistent with earlier measurements of AL charge in the presence of AOT~\cite{Guo2013}. The particles are positively charged at low AOT concentrations and get neutralized at concentration between 0.1 and 10~mM, where even a slight charge inversion is observed. At concentrations above 10~mM the particles get again slightly positively charged. Compared to the silica and SL particles the magnitude of the charge of the AL particles is smaller. For AL the maximal number of charges reached is about two times smaller than the maximal values for the other two surfaces. It seems that adsorption of charged micelles or surface groups solubilization is much less pronounced for AL surfaces.

\subsection{Interactions between Particles}

\begin{figure}[t]
\centering
\includegraphics[width=8.5cm]{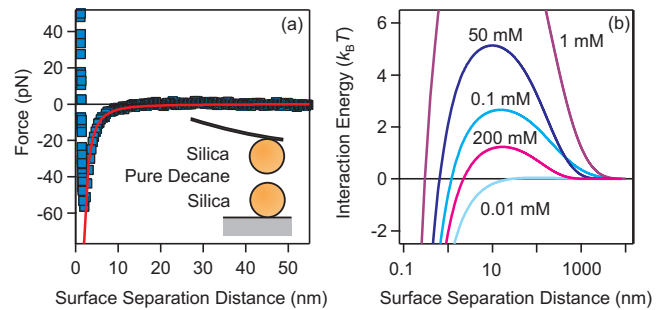}
\caption{(a) Van der Waals interaction measured between two $4~\mu$m silica spheres measured in pure decane. The line represents the fit to non-retarded van der Waals expression with Hamaker constant of $H = (1.8\pm 0.2)\cdot10^{-21}$~J. (b) DLVO interaction energies for 200~nm silica particles in decane for different AOT concentrations, calculated with~\eq~\ref{eq:dlvo}. The measured Hamaker constant was used to calculate van der Waals interaction, while electrokinetic potential and conductivity measurements were used to calculate the double-layer interaction.}
\label{fig:interactions}
\end{figure}

In order to understand the aggregation mechanisms in decane/AOT suspensions, we next focus on the particle interactions in these systems. The interactions are modelled with the DLVO theory as a sum of van der Waals, $U_{\rm vdW}$, and double-layer, $U_{\rm dl}$, interaction energies:
\begin{equation}
U_{\rm dlvo} = U_{\rm vdW} + U_{\rm dl} .
\label{eq:dlvo}
\end{equation}
The van der Waals interaction is calculated using the non-retarded expression for two spherical particles~\cite{Elimelech2013,Israelachvili2011},
\begin{equation}
U_{\rm vdW} = -\frac{H}{6}\left [ \frac{2a^2}{r^2-4a^2} + \frac{2a^2}{r^2} + \ln\frac{r^2-4a^2}{r^2} \right] ,
\label{eq:vdw}
\end{equation}
where $H$ is the Hamaker constant, and $r$ is the particle center-to-center distance. For the case of silica particles we have measured the van der Waals force between two micron-sized spheres with the colloidal probe technique, see Fig.~\ref{fig:interactions}a.
%%%%%%%%%%%%%%%%%%%%%%%%%%%%%
%%%%%%%%%%%%%%%%%%%%%%%%%%%%%
We have fitted the force curve with the expression $F_{\rm vdW} = -Ha/(12h^2)$~\cite{Elimelech2013}, where $h = r-2a$ is the surface separation distance. Note that this expression can be derived from \eq~\ref{eq:vdw} for $a\ll h$ and it is valid for the large 4~$\mu$m silica particles. The force measurements enabled us to extract the Hamaker constant, $H = (1.8\pm 0.2)\cdot10^{-21}$~J, for two silica surfaces interacting through decane. This constant is slightly smaller as compared to the value measured for similar silica particles across water~\cite{Uzelac2017} and close to the calculated value of two interacting silica surfaces across dodecane~\cite{Israelachvili2011}. The Hamaker constant extracted from the force measurements was later used to estimate the van der Waals interaction energy between aggregating silica particles in suspensions.

The other contribution to the DLVO interaction is the double-layer interaction energy, which is approximated by the Yukawa potential
\begin{equation}
U_{\rm dl} =  \frac{(e_0\zeta^*)^2}{\kT}\cdot\frac{a^2}{\lambda_{\rm B}}\cdot\frac{e^{-\kappa^* h}}{r},
\label{eq:dl}
\end{equation}
where $\zeta^*$and $\kappa^*$ are effective surface potential and inverse screening length, respectively. Measurements of electrostatic interactions in nonpolar solvents, have shown that the Yukawa approximation is highly accurate at large separation distances~\cite{Espinosa2010, Hsu2005, Sainis2007a, Sainis2008, Finlayson2016a}, where it is possible to measure double-layer force in these systems. The effective parameters entering (\ref{eq:dl}) are not equal to the true surface potential and to the inverse Debye length in general due to the possible non-linearity of electrostatics in suspensions at high surface charge and due to finite volume fraction effects. Recently, it has also been suggested that $\kappa^*$ can differ from the Debye length at high concentrations of salt in non-polar solvents as a consequence of charge regulation effects~\cite{Waggett2018}. The latter are also important for particle aggregation as will be discussed hereafter.

In order to make an estimation of the double-layer forces, $\zeta^*$ and $\kappa^*$ have to be estimated . The screening lengths estimated from electrical conductivity usually match well the decay lengths determined directly from experimental pair interaction curves~\cite{Hsu2005, Espinosa2010, Sainis2008}. Indeed, such experiments are conducted with two particles in an otherwise empty electrolyte whereas $\kappa^*$ essentially differs from the inverse Debye length at finite volume fractions. Therefore we have used values of screening length shown in Fig.~\ref{fig:conductivity}b for our calculations. Note that the Yukawa potential form {(\ref{eq:dl})} has been demonstrated only for weakly charged, well separated, colloids as a result of the DH theory. However, the electric field generated by highly charged colloids decays away from their surfaces so that at large enough distance it can be matched by a fictitious field obeying the DH equation and that would be generated by colloids with a so-called renormalized surface potential $\zeta^*$. This fitting procedure is termed renormalization method and can be undertaken for example with the recipe proposed by Trizac et al.~\cite{Trizac2003a}. In these conditions, it is not unreasonable to assume that colloids interact with a potential with the form ({\ref{eq:dl}}) provided the renormalized potential value is used. Experiments reveal that the electrokinetic, or zeta, potential value is usually quite consistent with the renormalized potential value, probably because using the  H\"uckel formula (\ref{eq:huckel}) to convert the true mobility into a potential is akin to a renormalization method. Therefore we have used directly the electrokinetic potentials, $\zeta$, in \eq~{\ref{eq:dl}} for the calculations of the DLVO potential.

The DLVO pair interactions for 200~nm silica particles in decane/AOT solutions at several AOT concentrations are shown in Fig.~\ref{fig:interactions}b. At 0.01 mM AOT the charge of the silica is practically zero and only the van der Waals force is present. The energy barrier increases for 0.1 and 1~mM AOT as the charge of the particles increases. At larger concentrations of 50 and 200~mM, the re-entrant behavior is observed as the energy barrier gets lowered again. Interactions are dominated by the charging of the silica surfaces.

The interactions for silica, SL, and AL particles were calculated in the full concentration range and were used to predict the stability of particle suspensions, presented below.

\subsection{Particle Aggregation}

Finally, we focus on particle aggregation in decane/AOT solutions. We have measured the kinetics of aggregation with light scattering and extracted the aggregation rate constant. We have further determined the absolute aggregation rates for rapidly aggregating suspensions in both water and decane, which are given in Table~\ref{tab:particles}. These fast rates are measured at conditions where the repulsive electrostatic forces are weak and attractive van der Waals interactions are dominant. For aqueous suspensions, this regime can be achieved by adding a sufficient amount of monovalent salt (KCl), which screens the electrostatic interactions. In the case of decane suspension we determine the fast rate at an AOT concentration for which the charge of the particles is close to zero. The fast rates in water are comparable to the values measured earlier for similar systems~\cite{Kobayashi2005, Oncsik2016, Sugimoto2018}, albeit we measure slightly lower rates for silica suspension. The fast rates measured in decane solutions are also comparable to the values obtained in water, see Table~\ref{tab:particles}. The viscosity of decane is slightly lower than that of water so the rates in decane should be a few percent higher, however this difference is small compared to the measurement errors. The fast rates in decane for latex particles are slightly lower than those obtained in water, and this difference could be related to weaker van der Waals forces in decane solutions, due to smaller difference in the refractive index of decane and particles.

\begin{figure}[t]
\centering
\includegraphics[width=6.0cm]{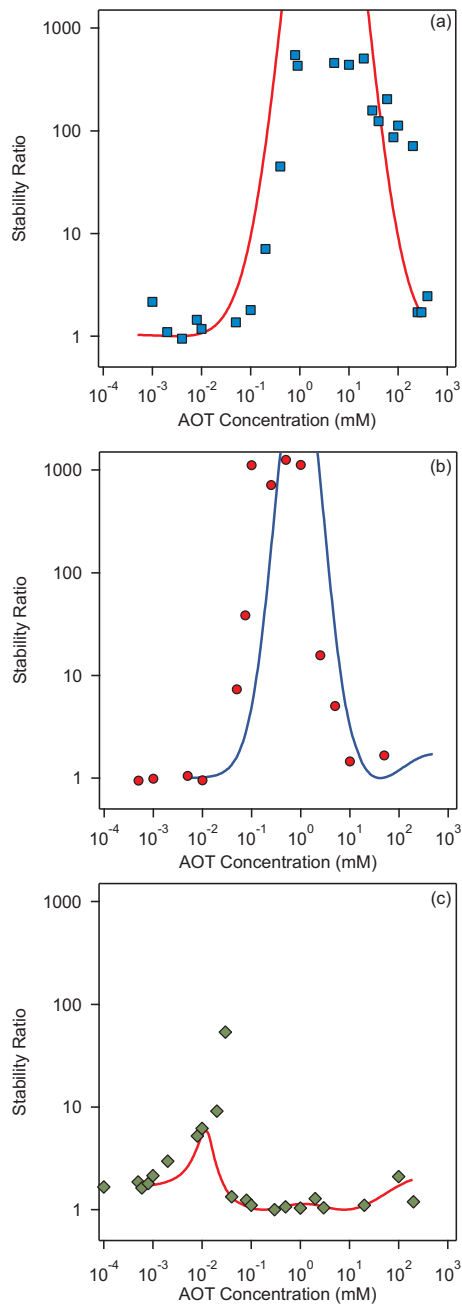}
\caption{Stability ratios in decane as a function of AOT concentration for (a) silica, (b) SL, and (c) AL particles. The symbols are measurements, while lines present model calculations based on DLVO interactions.}
\label{fig:stability}
\end{figure}

The apparent dynamic rate constants measured in decane for the wide concentration range of AOT were converted into the stability ratios via \eq~\ref{eq:stab_ratio}. The stability ratio, $W$, is a measure of the stability of suspensions, with $W = 1$ corresponding to fast aggregation where only attractive forces are presents, and with large values of $W$ corresponding to slow aggregating due to repulsive interactions. The stability ratios in decane as a function of AOT concentration for silica, SL, and AL particles are shown in Fig.~\ref{fig:stability}.
%%%%%%%%%%%%%%%%%%%%%%%%%%%%%
%%%%%%%%%%%%%%%%%%%%%%%%%%%%%
All three systems exhibit a similar qualitative behavior. The stability ratio is equal to unity at low AOT concentrations, the suspensions are not stable. At intermediate concentrations a peak in stability ratio is observed corresponding to stable dispersions. The stability again decreases at high concentrations and the suspensions start to aggregate fast again. While silica and SL reach high stability with $W\sim 1000$ at intermediate concentrations, the AL particles only reach maximal stability ratios between 10 and 100. This lower stability of AL particles can be explained by lower magnitude of surface charge of AL as compared to the other two systems. The qualitative stability behavior of all three suspensions can be understood by looking at the particle charging represented in Fig.~\ref{fig:mobility}. At low AOT concentration particles are not charged and suspensions are not stable, at intermediate concentration the magnitude of the surface charge increases and thus stability ratios increases, at high concentrations the particles get again neutralized and this process leads to unstable suspensions.

A better understanding of aggregation can be gained by calculating the stability ratios from particle interactions. To that end we have used DLVO interactions to model the aggregation behavior. The aggregation rate constants can be calculated by solving the diffusion equation for particles interacting via pair potential~\cite{Elimelech2013, Russel1989}
\begin{equation}
k = \frac{4\kT}{3\eta a} \left [ \int_0^\infty \frac{B(h/a)}{(2a+h)^2} e^{U_{\rm dlvo}/(\kT)} \right ]^{-1} ,
\label{eq:calc_rate}
\end{equation}
where $B(h/a)$ is a hydrodynamic resistance function approximated by~\cite{Honig1971}
\begin{equation}
B(h/a) = \frac{6(h/a)^2+13(h/a)+2}{6(h/a)^2 + 4(h/a)} .
\label{eq:hydro_resistance}
\end{equation}
The stability ratio is then calculated with $W = k_{\rm fast}/k$, where $k_{\rm fast}$ is calculated with~\eq~\ref{eq:calc_rate} and by setting $U_{\rm dl}$ to zero. The calculations of stability ratios for all three systems were first performed by utilizing the DLVO interaction energy given by (\ref{eq:dlvo}), (\ref{eq:vdw}) and (\ref{eq:dl}). This is justified by the use of effective parameters in (\ref{eq:dl}) and by the low volume fraction ($<4\cdot 10^{-5}$) of all the suspensions considered.

For calculating van der Waals interaction energy values of Hamaker constant of $1.8\cdot 10^{-21}$~J and $3\cdot 10^{-21}$~J were used for silica and polystyrene particles, respectively. Note that the first value was measured with direct force measurement, while the second was chosen based on values measured for SL and AL particles in water. The theoretical values of Hamaker constant for latex particles interacting across water are $9-13\cdot 10^{-21}$~J~\cite{Israelachvili2011}. However, experimentally substantially lower values $2-4\cdot 10^{-21}$~J have been measured, due to roughness effects~\cite{Moazzami-Gudarzi2016a, Elzbieciak-Wodka2014}. The Hamaker constant for latex particles interaction across decane should be lower than its value for an aqueous suspension and higher than  the silica-decane-silica value~\cite{Israelachvili2011}. Therefore the value of $3\cdot 10^{-21}$~J was chosen for polystyrene-decane-polystyrene system. Furthermore, the precise choice of Hamaker constant does not substantially affect the calculated stability ratios as it will be shown below.

The double-layer interaction energies were calculated by employing the interpolated values of electrokinetic potentials shown in Fig.~\ref{fig:mobility} and by inserting them into \eq~\ref{eq:dl}.

The stability ratios calculated by employing DLVO interaction energies are shown alongside experimentally measured values in Fig.~\ref{fig:stability}. The calculations can reproduce the observed behavior surprisingly well. There are slight shifts on the concentration axis and the peak magnitude of the stability ratio is not well predicted but this simple model is able to predict quite quantitatively the AOT concentration range in which the suspension is effectively stabilized. This reasonable agreement could be considered as  rather surprising, since some recent results suggested that the Yukawa model should fail in nonpolar suspensions~\cite{Hallett2018,  Waggett2018, Smith2017b}. We will come back to this apparent discrepancies later. From our results on aggregation one can conclude that DLVO forces are the principle drivers of aggregation in the decane/AOT solutions.

\begin{figure}[t]
\centering
\includegraphics[width=8.5cm]{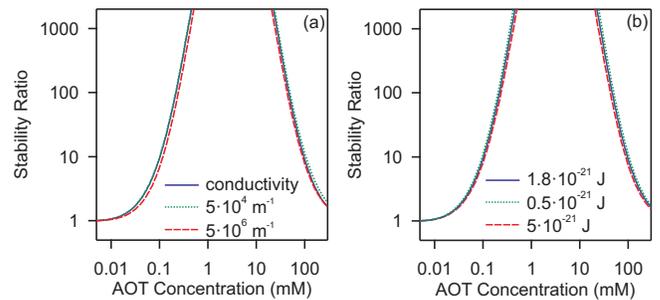}
\caption{Calculated stability ratios for silica in decane as a function of AOT concentration. (a) Effect of inverse Debye length on stability ratio. (b) Effect of Hamaker constant on stability ratio. In (a) solid line represents values extracted from conductivity shown in Fig.~\ref{fig:conductivity}b, and dotted and dashed lines use the constant values of $\kappa = 5\cdot 10^4$~m$^{-1}$ and $\kappa = 5\cdot 10^6$~m$^{-1}$, respectively. In (b) solid line represents results for experimentally measured Hamaker constant of $1.8\cdot 10^{-21}$~J, and the dotted and dashed lines use values of $H = 0.5\cdot 10^{-21}$~J and $H = 5\cdot 10^{-21}$~J.  }
\label{fig:stability_vdw}
\end{figure}

Let us now discuss the most important factors affecting the stability of nonpolar suspensions. We will focus on a silica system. In Fig.~\ref{fig:stability_vdw} we show the stability ratios, where we vary the inverse Debye length and the Hamaker constant in the calculation.
%%%%%%%%%%%%%%%%%%%%%%%%%%%%%
%%%%%%%%%%%%%%%%%%%%%%%%%%%%%
For the inverse Debye length we show three cases. In the first one we use the $\kappa$ determined experimentally for the calculation, the second represents the lower bound, and the third the upper bound as determined from the conductivity, see Fig.~\ref{fig:conductivity}. Note that in the second and third case we fix $\kappa$ and do not change it as a function of the concentration. Interestingly, the calculated stability ratios are practically independent on the choice of the screening length. This behavior is different compared to the aqueous systems, where the screening length is an extremely important factor for determining the stability~\cite{Cao2017, Cao2018}. This insensitivity to the screening length shows that AOT influences aggregation by acting on the surface charge density rather than by setting the screening length of the double-layer. This is not surprising since all the systems investigated (whatever the AOT concentration) are in the $\kappa a\ll 1$ limit for which electrostatics converge to those of the salt-free regime where the only ions in the suspension are the counterions of the colloids. This observation is also consistent with electrokinetic measurements, which show that the change in the electrokinetic potential is caused by the change of the charge and not by the double-layer screening, see Fig.~\ref{fig:mobility}.

Next, we focus on the effect of the van der Waals forces on the aggregation. Although attractive van der Waals forces cause the particles to aggregate, the choice of the value of the Hamaker constant does not significantly affect the stability ratio, as shown in Fig.~\ref{fig:stability_vdw}b. By changing Hamaker constant from 0.5 to $5\cdot 10^{-21}$~J the stability ratio is practically unaffected.

\begin{figure}[t]
\centering
\includegraphics[width=8.5cm]{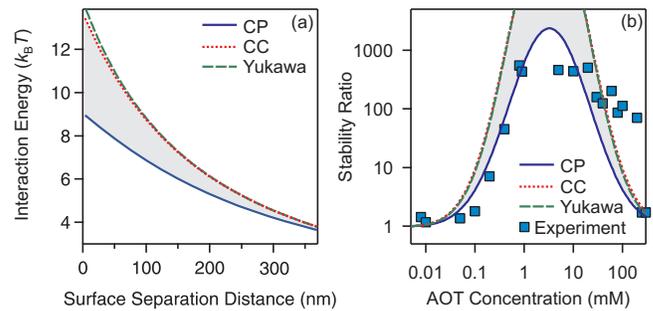}
\caption{Effect of charge regulation on interactions and stability ratios of silica in decane as a function of AOT concentration. (a) Interaction energy calculated with constant potential (CP) and constant charge (CC) boundary conditions. Yukawa interaction potential is added for comparison. The parameters correspond to AOT concentration of 4.7 mM. (b) Stability ratios calculated using CP, CC, and Yukawa interaction potentials. The area between CP and CC is shaded. Experimental results are shown with blue squares.}
\label{fig:stability_regulation}
\end{figure}

Let us finally focus on the influence of charge regulation effects on aggregation in the present system. Here we refer to charge regulation as the variation of charge as a function of the distance between two interacting particles~\cite{Trefalt2016a, Adzic2014}. In addition, the charge of the particles can also vary with salt or particle concentration, and pH and the two phenomena are closely related~\cite{Trefalt2016a,Podgornik2018, Avni2019}. While charge regulation in aqueous media has been thoroughly studied~\cite{Trefalt2016a, Avni2019}, recent measurements also showed the importance of charge regulation in nonpolar suspensions~\cite{Hallett2018}. 
It is important to realize that charge regulation effects cannot be accounted for with the Yukawa potential (\ref{eq:dl}) for a constant $\zeta^*$ because in this case the same numerical value of the prefactor of the exponential can result from a constant charge condition or a constant potential condition, among other more sophisticated conditions. In order to address the influence of charge regulation on the stability of colloidal particles in decane/AOT we compare the stability ratios based on exact interaction potentials stemming from constant potential (CP) and constant charge (CC) boundary conditions used to compute the exact solution of the Debye-H\"uckel equation for a pair of spheres~\cite{Carnie1993}, see Fig.~\ref{fig:stability_regulation}.
%%%%%%%%%%%%%%%%%%%%%%%%%%%%%
%%%%%%%%%%%%%%%%%%%%%%%%%%%%%
The interaction energies between two silica particles at conditions corresponding to 4.7~mM AOT are shown in Fig.~\ref{fig:stability_regulation}a.  They show that the CC matches the Yukawa profile, which is expected for small $\kappa a$~\cite{Carnie1993, Glendinning1983}. The CP curve is below CC and Yukawa and the three curves converge to the same value at large separations. The stability ratios calculated with CC and CP boundary conditions are presented in Fig.~\ref{fig:stability_regulation}b. Experimental values and Yukawa calculations are also added for comparison. The CC and Yukawa predict stronger stability as compared to the CP solution, and such behavior is expected from the corresponding interaction potentials. Most interestingly, the CP curve matches very nicely with experimental measurements. This result suggests that silica surfaces regulate strongly and get progressively neutralized upon approach. The adjustment of the charge upon approach is regulated by adsorption/desorption equilibria, which involves AOT surfactant in a peculiar way that seems to maintain a roughly constant surface potential. Recent electrophoretic mobility measurements have also shown strong charge regulation upon increasing particle concentration for polymeric colloids in AOT/dodecane suspensions~\cite{Hallett2018} and in pure isopropanol~\cite{Teulon2019}. It seems that for both colloidal stability and charging of particles at elevated concentrations charge regulation is driven by similar adsorption/desorption mechanism.

Let us finally discuss why the DLVO theory is quite appropriate for describing stability of particles in the present decane/AOT system. We have used the linear DH approximation for calculating double-layer interactions, with electrokinetic potentials considered as some kind of effective potentials. This is a classical and reasonable method {\em per se} that allows the mapping of true non-linear electrostatics due to high surface potentials on a linear theory suited for low potentials. However, we have also used the charge renormalization procedure introduced by Trizac et al.~\cite{Trizac2003a} to estimate what would be the true surface potential $\psi_s$ that would lead to the effective, measured, electrokinetic potential $\zeta^*$. The maximum difference was never larger than a few percent, showing that in the present case charges were always small enough to permit DH linearization. Further, concentration of particles in our experiments is low enough to avoid many-body forces~\cite{Merrill2009} and the non-monotonic dependence of surface charge at high particle concentrations described by Hallett et al.~\cite{Hallett2018}. Additionally, our results in Fig.~\ref{fig:stability_vdw}a show that the stability does not depend on the value of screening length. Therefore, our stability ratio results do not contradict recent findings of discrepancies between measured and predicted screening lengths in nonpolar media~\cite{Waggett2018}, but the aggregation behavior is simply not sensitive to the changes in the screening length.

Although the DLVO predictions for our three systems shown in Fig.~\ref{fig:stability} are in general good, the remaining discrepancies are probably due to charge regulation effects. Indeed, the surface charge variation upon approach of two particles affects the stability ratios as shown in Fig.~\ref{fig:stability_regulation}a with the use of potentials extracted from the exact DH solution. Further inconsistencies between theory and experiment could be connected to non-uniformity of surface charge. Since the surface charge is at most on the order of 10 charges per particle (Fig.~\ref{fig:mobility}), the distance between the charges on the particle surface is large and therefore charged patches exist. Such non-uniformity of charge can lead to additional non-DLVO forces~\cite{Adar2017, Adzic2015}. At high AOT concentrations the stability of the silica is underestimated by the theory. This discrepancy could be caused by steric effects due to adsorbed surfactant at high concentrations~\cite{Szilagyi2014a}.

\section{Conclusions}

We have measured the stability of silica and polystyrene latex particles in decane/AOT suspensions. Upon addition of AOT surfactant particles get charged and the suspensions become stable in the intermediate surfactant concentration range. At higher concentrations of AOT the particles are neutralized and the suspensions aggregate fast again. The DLVO theory based on a simple Yukawa potential is able to predict the surfactant concentration range in which the suspension is stabilized and to recover the re-entrant behavior observed experimentally. The aggregation behavior is driven by surface charging through adsorption/desorption processes, while screening of the double-layer does not play an important role in destabilization. On the other hand, charge regulation effects are an important factor affecting the value of the stability ratio. The silica particles in the decane/AOT system follow the model predictions calculated by using constant potential boundary conditions. As yet, stability has to be calculated with potentials extracted from the exact DH solution since no simple Yukawa-like model accounting for charge regulation and suited for the $\kappa a\ll1$ case is known. Building such a model would be highly valuable for applications.

\section*{Acknowledgements}
This research was supported by the Swiss National Science Foundation through grant 162420 and the University of Geneva. The authors are thankful to Michal Borkovec for providing access to the instruments in his laboratory and to Plinio Maroni for the help with AFM measurements. M.F also acknowledges the support of Iran Ministry of Science, Research and Technology.

\bibliography{paperLib}

\bibliographystyle{apsrev4-1}

\end{document}